\documentclass[conference]{IEEEtran}
\usepackage[utf8]{inputenc}
\usepackage{times,graphics, dsfont, epsfig,amsmath,xspace,endnotes,pifont,multirow,rotating,listings,amssymb,algorithmic,color,caption,nicefrac,adjustbox,tabularx,mathtools, algorithmic,algorithm}
\usepackage{graphicx}
\usepackage{tabularx}
\newcolumntype{L}[1]{>{\raggedright\arraybackslash}p{#1}} 
\newcolumntype{C}[1]{>{\centering\arraybackslash}p{#1}} 
\newcolumntype{R}[1]{>{\raggedleft\arraybackslash}p{#1}} %

\usepackage{steinmetz}
\usepackage{amsmath}
\usepackage{amssymb}
\usepackage{float}
\ifCLASSOPTIONcompsoc
    \usepackage[caption=false, font=normalsize, labelfont=sf, textfont=sf]{subfig}
\else
\usepackage[caption=false, font=footnotesize]{subfig}
\fi
\usepackage{graphicx}
\usepackage{hyperref}
\usepackage{url}
\usepackage{tikz}

\usepackage[letterpaper, left=0.625in, right=0.625in, bottom=1in, top=0.75in]{geometry}
\usepackage{amsthm}
\newtheorem{thm}{Theorem}

\newtheorem{problem}[thm]{Problem}

\newcommand{\A} {{\bf{A}}}

\newcommand{\F} {{\bf{F}}}
\newcommand{\D} {{\bf{D}}}
\newcommand{\I} {{\bf{I}}}

\renewcommand{\c} {{\bf{c}}}
\newcommand{\g} {{\bf{g}}}
\newcommand{\f} {{\bf{f}}}
\renewcommand{\v} {{\bf{v}}}

\newcommand{\h} {{\bf{h}}}

\renewcommand{\d} {{\bf{d}}}

\author{
\IEEEauthorblockN{Nariman Torkzaban}
\IEEEauthorblockA{University of Maryland,
College Park\\
narimant@umd.edu}
\and
\IEEEauthorblockN{Mohammad A. (Amir) Khojastepour}
\IEEEauthorblockA{NEC Laboratories,
America\\
amir@nec-labs.com}
\and
\IEEEauthorblockN{John S. Baras}
\IEEEauthorblockA{University of Maryland,
College Park\\
baras@umd.edu}
}

\def\BibTeX{{\rm B\kern-.05em{\sc i\kern-.025em b}\kern-.08em
    T\kern-.1667em\lower.7ex\hbox{E}\kern-.125emX}}

\newcommand\copyrighttext{%
  \footnotesize \textcopyright 2022  IEEE. Personal use of this material is permitted.
  Permission from IEEE must be obtained for all other uses, in any current or future
  media, including reprinting/republishing this material for advertising or promotional
  purposes, creating new collective works, for resale or redistribution to servers or
  lists, or reuse of any copyrighted component of this work in other works.
  }
\newcommand\copyrightnotice{%
\begin{tikzpicture}[remember picture,overlay]
\node[anchor=south,yshift=10pt] at (current page.south) {\fbox{\parbox{\dimexpr\textwidth-\fboxsep-\fboxrule\relax}{\copyrighttext}}};
\end{tikzpicture}%
}

\begin{document}
%
% paper title
% Titles are generally capitalized except for words such as a, an, and, as,
% at, but, by, for, in, nor, of, on, or, the, to and up, which are usually
% not capitalized unless they are the first or last word of the title.
% Linebreaks \\ can be used within to get better formatting as desired.
% Do not put math or special symbols in the title.
\title{Codebook Design for Composite Beamforming in Next-generation mmWave Systems}
%\title{Virtualized Network Function Placement\\for Cellular Network Slicing}

% author names and affiliations
% use a multiple column layout for up to three different
% affiliations

% use for special paper notices
%\IEEEspecialpapernotice{(Invited Paper)}

% make the title area
\maketitle
\copyrightnotice
% As a general rule, do not put math, special symbols or citations
% in the abstract
\begin{abstract}
In pursuance of the unused spectrum in higher frequencies, millimeter wave (mmWave) bands have a pivotal role. However, the high path-loss and poor scattering associated with mmWave communications highlight the necessity of employing effective beamforming techniques. 
In order to efficiently search for the beam to serve a user and to jointly serve multiple users it is often required to use a \emph{composite beam} which consists of multiple disjoint lobes.
A composite beam covers multiple desired angular coverage intervals (ACIs) and ideally has maximum and uniform gain (smoothness) within each desired ACI, negligible gain (leakage) outside the desired ACIs, and sharp edges. We propose an algorithm for designing such ideal composite codebook by providing an analytical closed-form solution with low computational complexity. 
There is a fundamental trade-off between the gain, leakage and smoothness of the beams. Our design allows to achieve different values in such trade-off based on changing the design parameters.
We highlight the shortcomings of the uniform linear arrays (ULAs) in building arbitrary composite beams. Consequently, we use a recently introduced twin-ULA (TULA) antenna structure to effectively resolve these inefficiencies. Numerical results are used to validate the theoretical findings.
% With the ever-increasing desire to explore higher bandwidths in search for abundance of unused spectrum, millimeter wave (mmWave) communications has attracted a lot of attention from both the academia and industry. 
\end{abstract}

% no keywords

% For peer review papers, you can put extra information on the cover
% page as needed:
% \ifCLASSOPTIONpeerreview
% \begin{center} \bfseries EDICS Category: 3-BBND \end{center}
% \fi
%
% For peerreview papers, this IEEEtran command inserts a page break and
% creates the second title. It will be ignored for other modes.
\IEEEpeerreviewmaketitle

\section{Introduction}

%\amir{Emphasize on the importance and use cases of composite beamforming. Cirrent version only addresses the beamforming not composite.}
%\nariman{need to reflect the importance of composite beamforming }
With the exponential growth in the number of users and the diversity of the broadband applications in next-generation communication systems, the ever-increasing need to explore higher bandwidths, reveals the pivotal role of mmWave communications in future wireless networks. However, given the high path-loss and poor scattering associated with mmWave communications, effective beamforming techniques integrating a large number of antennas are required to ensure satisfactory quality of service (QoS).

Due to the large scale of massive MIMO systems and consequently the large mmWave channel matrix, the full channel state information (CSI) %required for constructing effective beams are 
is hardly available to the transmitter. Therefore, feedback-based beamforming techniques 
%based on codebook design need to be 
are employed for efficient mmWave communications. On the other hand, the cost of having one RF chain per each transmit antenna is prohibitive. Hence, not only the multitude of RF chains for massive MIMO systems incur a considerable cost and power consumption, but also it is not going to be used to achieve multiplexing gain as the full CSI is not practically available. Therefore more practical system designs consider reducing the number of RF chains to achieve lower power consumption and lower cost. Such designs that only employ a single RF chain are denoted as analog beamforming structures in the literature, while hybrid beamforming is reserved for the designs consisting of a few RF chains. In hybrid beamforming, the beamformer  consists of two layers. At the first layer, the baseband beamforming unit (BBU) performs digital beamforming, i.e. controls both the gain and the phase of the input symbol, while at the second layer, the radio remote head (RRH) only performs phase shifts, i.e. realizes the analog beamforming. 

Due to the nature of mmWave channels as being largely line of sight and having only a few dominant paths, physical beamforming (beam steering) is a practical an effective way. The communication beams are designed to have maximum gain toward the direction of the angle of departure of the user channel. Often, a beam search procedure, e.g., sequential beam search \cite{nitsche2015steering}, hierarchical beam search \cite{noh17}, interactive \cite{khal20} and non-interactive beam search \cite{SS19}, is used to find the best communication beam. The proposed algorithms in the literature often ignores the effect of multipath. % or finds a remedy to deal with multipath in the beam search. 
Recent works \cite{shah20} have shown that it is in fact possible to exploit multipath to our benefit in order to find more robust beams that are less susceptible to blockage and shadowing. The key idea in \cite{shah20} is to design a \emph{composite beam} that has multiple lobes that is covering the dominant paths of the user channel.

%In such methods, each codebook consists of several codewords that are known to both the transmitter and the receiver. Each codeword a.k.a \emph{beamforming array} is an array of weights that uniquely determines a specific beam pattern. The feedback-based procedure outputs the beam that is preferred for communications by the receiver. The codebook design problem is at the core of an efficient beamforming scheme. There is a trade-off between the size of the codebook and the narrowness of the beams. Although narrower beams are preferred,the larger size of the associated codebook will render a more complicated feedback-based beamforming. Therefore, the codebook design problem is challenging and needs to be addressed properly. 

%We note that in massive MIMO systems the cost of deploying one radio-frequency (RF) chain per antenna is prohibitive. Therefore more practical system designs consider reducing the number of RF chains to achieve lower power consumption and lower cost. Such designs that only employ a single RF chain are denoted as analog beamforming structures in the literature, while hybrid beamforming is reserved for the designs consisting of a few RF chains. In hybrid beamforming, the beamformer  consists of two layers. At the first layer, the baseband beamforming unit (BBU) performs digital beamforming, i.e. controls both the gain and the phase of the input symbol, while at the second layer, the radio remote head (RRH) only performs phase shifts, i.e. realizes the analog beamforming. 

%\nariman{fixing adding or removing citations + use-cases of composite beamforming }
In this paper, we propose an algorithm to design such \emph{composite beam}, i.e., the beam that are comprised of multiple non-neighboring angular coverage intervals (ACIs), say in azimuth direction, of possibly different widths.
%the \emph{composite codebook} design problem for \emph{composite beamforming} in mmWave systems, whereby we are interested in devising codebooks that are supposed to cover \emph{composite beams}, i.e., beams that are comprised of multiple non-neighboring angular coverage intervals (ACIs) of possibly different widths. 
The composite beams are not only important as a data communication beams, they can also facilitate the beam search. A codebook of composite beams, \emph{composite codebook}, is designed for a set of composite beams that are defined over a set of desired ACIs \emph{ ACI set}. Each entry of the codebook is a beamforming vector that generates a composite beam defined as a beam which covers a union of disjoint ACIs out of a set of all desired ACIs. 
%More precisely, the codebook design problem takes as input the set of desired ACIs denoted by \emph{ ACI set}. Each subset of the ACI set specifies a composite beam that has to be covered by a \emph{composite codeword}. The composite codebook is the set of all composite codewords. 
Such composite codebook can be used in variety of applications in next-generation mmWave communications such as user tracking \cite{shah19}, target monitoring \cite{Nos19}, 5G positioning, two-way communications \cite{Ata20}, design of reconfigurable intelligent surfaces\cite{glo21}, UAV-enabled networks \cite{gholami2020joint}, etc. The composite codebook design problem may be also viewed as generalized version of the codebook design problem \cite{love15}\cite{noh17} where the angular range under study is divided into equal-length ACIs and each codeword is supposed to cover only a single ACI. 
%\nariman{maybe we want to add the main contributions bullet-wise here. }

%The main contributions of this paper are as follows: 

\textbf{Notations:} Throughout this paper, $\mathbb{C}$ denotes the set of complex numbers, $\mathcal{C N}\left(m, \sigma^{2}\right)$ denotes the complex normal distribution with mean $m$ and variance $\sigma^{2}$, $[a, b]$ is the closed interval between $a$ and $b, \mathbf{1}_{a, b}$ is the $a \times b$ all ones matrix, $\mathbf{I}_{N}$ is the $N \times N$ identity matrix, is the ceiling function, $\mathds{1}_{[a, b)}$ is the indicator function, $\|\cdot\|$ is the $2$ -norm, $|.|$ is the $1$-norm, $\odot$ is the Hadamard product, $\otimes$ is the Kronecker product, $\mathbf{A}^{H},$ and $\mathbf{A}_{a, b}$ denote conjugate transpose, and $(a, b)^{t h}$ entry of \A.

The remainder of the paper is organized as follows. Section~\ref{sec:desc} describes the system model. In Section~\ref{sec: formulation} we formulate the codebook design problem and propose our solutions in section~\ref{sec: proposed}. Section~\ref{sec:evaluation} presents our evaluation results, and finally, in Section~\ref{sec:conclusions}, we highlight our conclusions and discuss directions for future work.

\section{System Model } 
\label{sec:desc}

\subsection{Channel Model}

We consider a mmWave channel between a multi-antenna base station (BS) on the transmitter side and a single-antenna user equipment (UE) on the receiver side. The channel is defined as
\begin{equation}
y=\sqrt{\rho} \mathbf{h}^{H} \mathbf{c} s+n    
\end{equation}
where $\rho$ denotes the system signal-to-noise ration (SNR), $\h \in \mathbb{C}^{M_t}$ the block fading channel vector, $\c \in \mathbb{C}^{M_t}$ the unit-norm digital beamforming array ($\|\c\| = 1$), $s \in \mathbb{C}$ the input signal, and $n \sim \mathcal{C} \mathcal{N}(0,1)$, the additive white complex Gaussian noise.
Under the hybrid beamforming scheme, it holds that $\c = \F\v$, where  $\F=\left[\f_{1}, \cdots, \f_{N_{RF}}\right] \in \mathbb{C}^{M_{t} \times N_{RF}}$, where $\F$ is an analog beamsteering matrix only capable of phase shifting, where all the vectors $\f_n, n = 1\ldots N_{RF}$ are subject to the \emph{equal gain} constraint defined as $|\f_n^{(m)}| = 1, m = 1\ldots M_t$, and $\v \in \mathbb{C}^{N_{RF}}$ is the baseband beamforming vector. Similarly, the unit-norm constraint on $\c$, enforces that $\|\F\v\| = 1$.

\subsection{Beamforming Model}
We consider the design of physical beams that are steered in azimuth plane where the beams are supposed to cover one (or multiple disjoint) ACIs. An ACI covering the angular range from $\theta^s_b$ to $\theta^f_b$ is denoted by $\omega_b = [\theta^s_b, \theta^f_b)$ where $\theta^c_b = (\theta^s_b + \theta^f_b)/2$, and $\lambda_b = |\theta^s_b - \theta^f_b|$ are the center and beamwidth associated with the beam lobe covered by this ACI. We assume $\theta \in [-\pi,0]$. Further, let us introduce the change of variable $\psi = \pi cos\theta$. We have $\psi \in [-\pi,\pi]$, and each beam is represented by $\omega^{\psi}_b = [\psi^s_b, \psi^f_b)$ in the $\psi-\text{domain}$, where $\psi_b^a = \pi cos\theta_b^a$, $a \in \{s, f\}$. Further define $\delta_b = |\psi^s_b - \psi^f_b|$. For the rest of the paper, we prefer to work with the beams over the $\psi-\text{domain}$, unless otherwise stated.

Let $\mathcal{A}$ denote the set of ACIs defined for a given codebook design problem. Let each element of $\mathcal{A}$ be denoted by an index.
A beam is denoted by $\mathcal{B}(B) = \{w_b\}_{b \in B}$, where $B$ is the set of indices of non-neighboring ACIs. % that are covered by beam $\mathcal{B}(B)$. 
A set of ACIs is non-neighboring if any pair of ACIs in that set are not neighbors, i.e., the starting angle of one beam is not equal to the ending angle of the other beam. %union of any pair of ACIs cannot be represented as an ACI. 
A beam is called \emph{single beam} (\emph{composite beam}) if $|B| = 1$ ($|B| > 1$).

\begin{figure}
    \centering
    \subfloat[Angular Coverage Intervals (ACIs) \label{fig:aci}]{%4
            \includegraphics[width=0.5\linewidth]{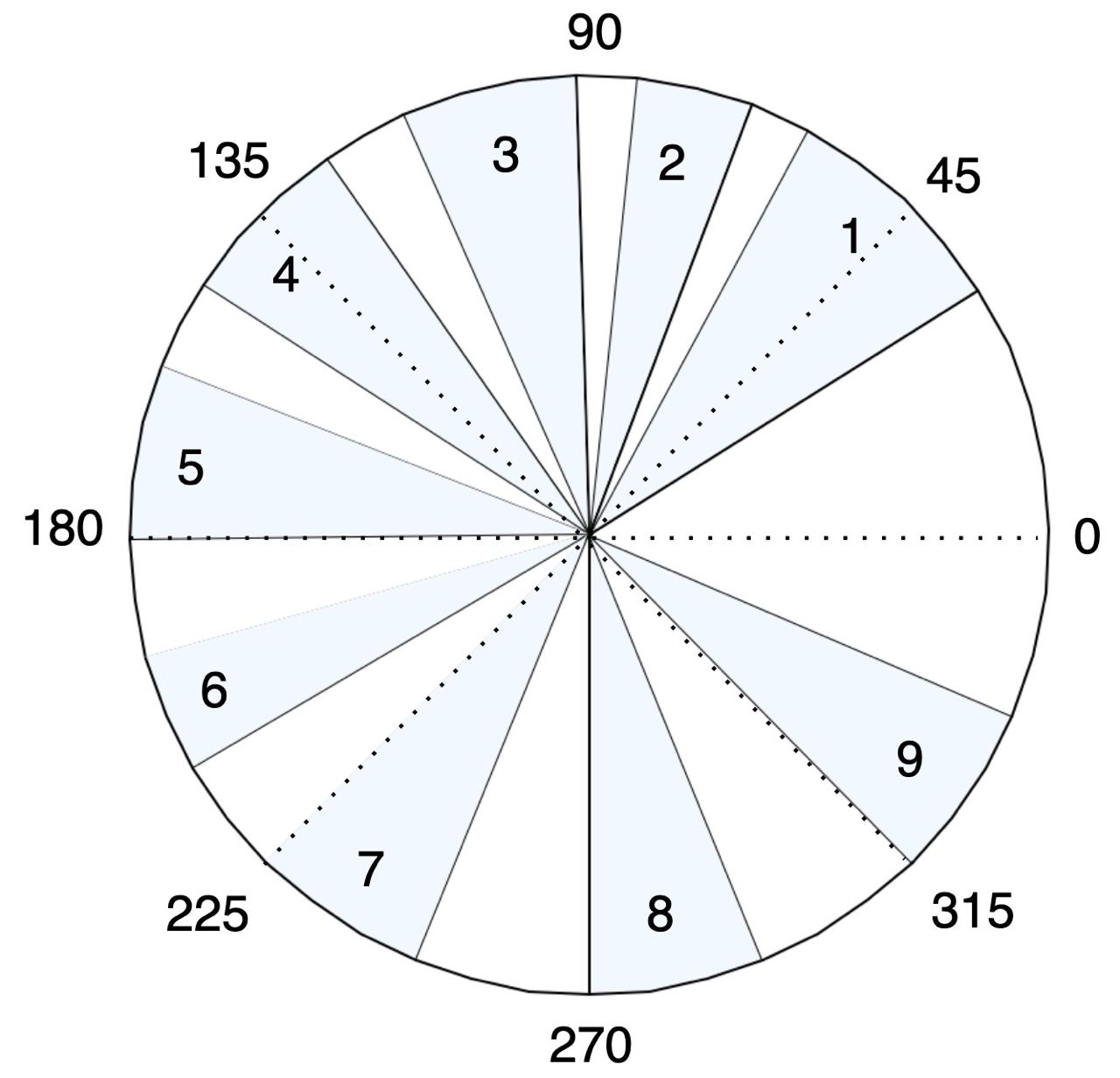}}
    \subfloat[ Two Composite Beams \label{fig:beam}]{%4
            \includegraphics[width=0.5\linewidth]{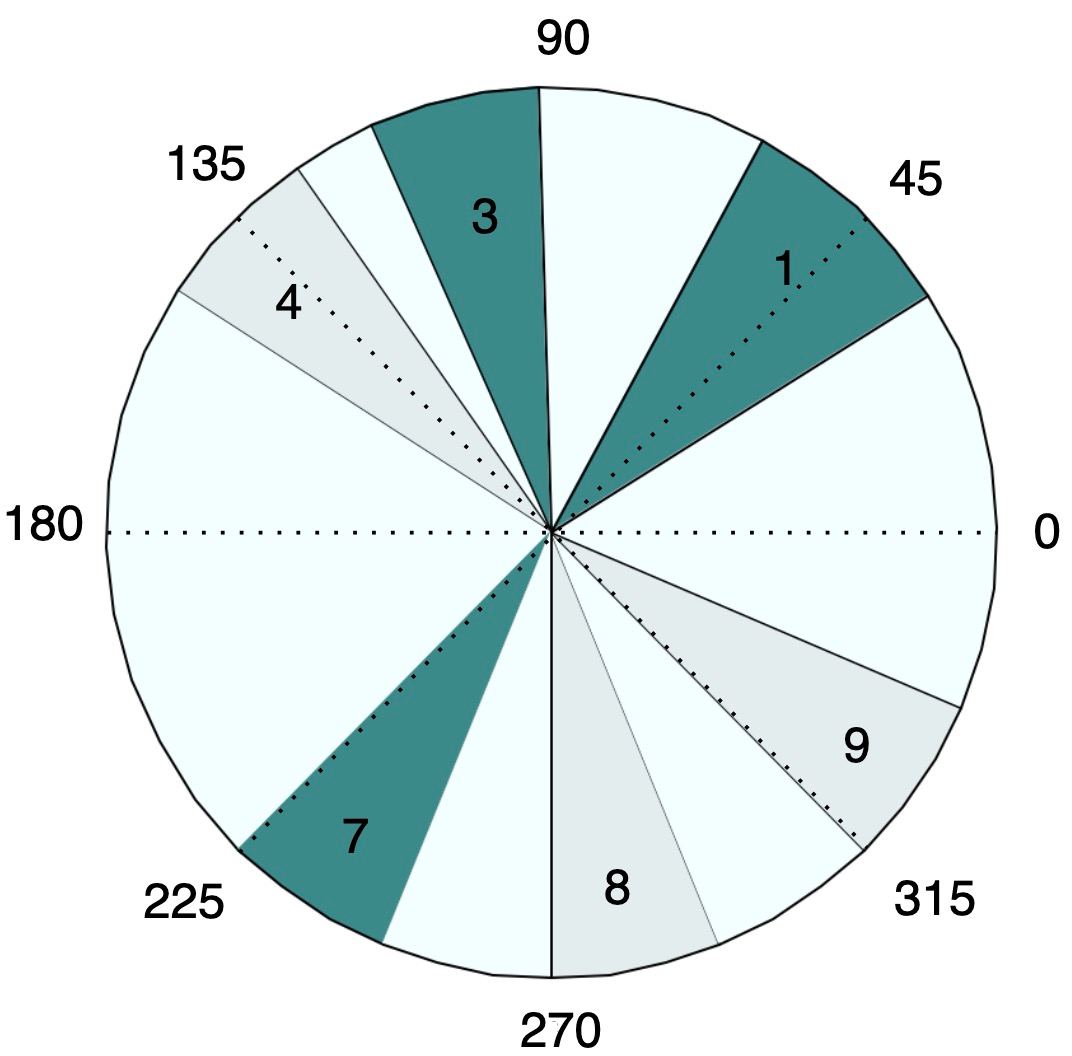}}
    \caption{Example of the Codebook Design Problem Settings}
    \label{fig:example}
\end{figure}

\noindent \textbf{Example 1.} One example of this setting is shown in fig.~\ref{fig:example}. Particularly, fig.~\ref{fig:aci} depicts a hypothetical ACI set given as the input to the codebook design problem. We have, 
\begin{align}
    \mathcal{A} &= \left\{[\frac{\pi}{6}, \frac{\pi}{3}), [\frac{19\pi}{48}, \frac{23\pi}{48}), [\frac{\pi}{2}, \frac{5\pi}{8}), [\frac{11\pi}{16}, \frac{13\pi}{16}), [\frac{7\pi}{8}, \pi),\right.\nonumber\\&\left. [\frac{-7\pi}{8}, \frac{-19\pi}{24}), [\frac{-3\pi}{4}, \frac{-5\pi}{8}), [\frac{-\pi}{2}, \frac{-3\pi}{8}), [\frac{-\pi}{4}, \frac{-\pi}{8}) \right\}\nonumber
\end{align}

Fig.~\ref{fig:beam} depicts two potentially desired composite beams $\mathcal{B}(B_1), \mathcal{B}(B_2) \subseteq \mathcal{A}$. We have, 
\begin{align}
    \mathcal{B}(B_1) &= \left\{[\frac{\pi}{6}, \frac{\pi}{3}),  [\frac{\pi}{2}, \frac{5\pi}{8}), [\frac{-3\pi}{4}, \frac{-5\pi}{8}) \right\}\nonumber
\end{align}
\begin{align}
    \mathcal{B}(B_2) &= \left\{[\frac{11\pi}{16}, \frac{13\pi}{16}),  [\frac{-\pi}{2}, \frac{-3\pi}{8}), [\frac{-\pi}{4}, \frac{-\pi}{8}) \right\}\nonumber
\end{align}
where, $B_1=\{1,3,7\}$, and $B_2=\{4,8,9\}$.

\noindent \textbf{Example 2.} One may be interested in designing a codebook where the beamwidth of each beam, whether single or composite, has a resolution of say $\pi/8$ and is not larger than $\pi/2$. In this case, the ACI set $\mathcal{A}$ consists of all ACIs like $\omega_b = [\theta^s_b, \theta^f_b)$ with beamwidth  $\lambda_b \in \{\pi/8, \pi/4, 3\pi/8, \pi/2\}$ and $\theta_b^s = k\pi/8$,  $k \in \mathbb{Z}_{\geq 0}$.

The first example shows the intuition behind our beamforming setting given an arbitrary ACI list. The second example highlights the case where the ACIs are overlapping.

Further, we say a vector $\g$  of length $L$ is a \emph{geometric vector} with parameter $\rho$ if and only if it can be written as
\begin{align}
   \g = [1, e^{j\rho}, \cdots, e^{j(L-1)\rho}]
\end{align}

\subsection{Antenna Array Model}

We consider two different antenna configurations, namely, uniform linear array (ULA), and twin-ULA (TULA) \cite{TULA}.
% \nariman{citing twin-ULA (TULA) milcom}. 
Under the ULA structure the antennas are placed uniformly along the $x-\text{axis}$, where each antenna has a distance of $d$ from its previous (next) antenna. However, under the TULA structure, the antennas are arranged in two  ULAs that are placed in parallel to the $x-\text{axis}$, and with a distance of $d_y$ along the $y-\text{axis}$. In this paper, we consider half-wavelength ULAs, i.e., $d = \frac{\lambda}{2}$. Moreover, under TULA, we set $d_y = \frac{\lambda}{3}$.  For each beamformer $\c$, under the ULA structure, the \emph{reference gain} $G^{ula}{( \psi, \c)}$ at every given direction $\psi$ is given by 
\begin{align}
    G^{ula}{( \psi, \c)} = \left| \d_{ula, M_t}^H(\psi)  \c \right|^2\label{reference gain}
\end{align}
\noindent where $\d_{ula, M_t}(\psi)$ is the array response vector (directivity) of the ULA in that direction and is given by, 
\begin{align}
    \d_{ula, M_t}(\psi) =  [1, e^{j{\psi }}, \ldots, e^{j{ (M_t-1)\psi}}]^T
    \label{array_factor}
\end{align}
Similarly, under the TULA structure it holds that, 
\begin{align}
    G^{tula}(\psi, \c) = \left| \d_{tula, M_t}^H  \c \right|^2 \label{tula_gain}
\end{align}
\noindent where the directivity of the TULA is given by,
\begin{align}
    &\d_{tula, M_{t}}(\psi, \phi) = \left[\d_{ula, \frac{M_{t}}{2}}^{T}(\psi), \quad e^{j\phi}\d_{ula, \frac{M_{t}}{2}}^{T}(\psi)\right]^T \label{tula_dir}
\end{align}
with $\phi = \frac{2\pi}{3}\sin\theta = \frac{2\pi}{3}\sqrt{1-\psi^2}$.
\begin{figure} 
    \centering
        \subfloat[ULA  \label{ULA_config}]{%5
            \includegraphics[width=0.4\linewidth]{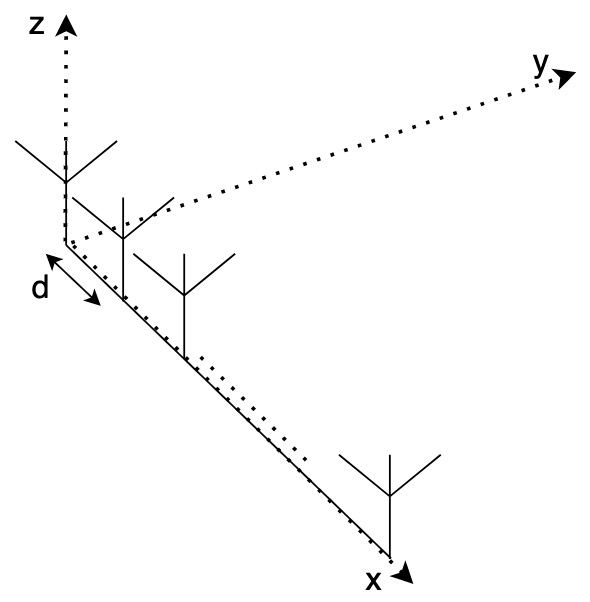}}
        \subfloat[TULA \label{TULA_config}]{%5
            \includegraphics[width=0.5\linewidth]{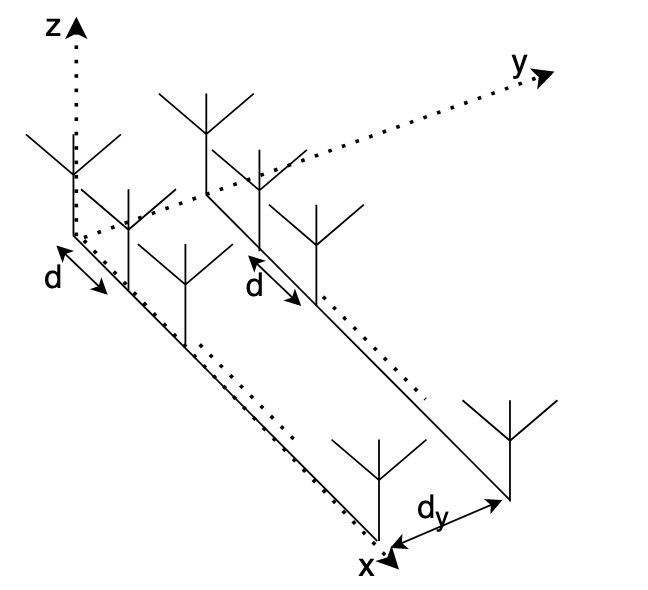}}
  \caption{Antenna Array Model}
  \label{fig:array_config} 
\end{figure}
In the next section, we propose the formulation for the codebook design problem.

\section{Composite Codebook Design Problem }
\label{sec: formulation}

Let $\c_B$ be the codeword corresponding to the beam $\mathcal{B}(B)$. Using the Parseval's theorem \cite{TULA}, it is straightforward to write, 
\begin{equation}
    \int_{-\pi}^{\pi} G(\psi, \mathbf{c}) d \psi=2 \pi\|\mathbf{c}\|^{2}=2 \pi
\end{equation}

Applying the last equation to the ideal gain corresponding to  the codeword $\c_B$ we get, 
\begin{align}
&\int_{-\pi}^{\pi} G_{\text {ideal }, B}(\psi) d \psi =\int_{\mathcal{B}(B)} t d \psi+\int_{[-\pi, \pi] \backslash \mathcal{B}(B)} 0 d \psi \nonumber\\
&= \sum_{b \in {B}}{\int_{\omega^{\psi}_{b}} t d \psi} = \sum_{b \in B}\delta_b t=2 \pi \label{composite}
\end{align}

Therefore, $t = \frac{2 \pi}{\Delta_B}$, where $\Delta_B = \sum_{b \in B}\delta_b$. It follows that, 
\begin{equation}
    G_{\text {ideal }, B}(\psi)=\frac{2 \pi}{\Delta_B} \mathds{1}_{\mathcal{B}^\psi(B)}(\psi), \quad \psi \in[-\pi, \pi] \label{composite_ideal_gain}
\end{equation}

We wish to find the optimal configuration $\c_B$ for the antenna array such that $G(\psi, \c)$ becomes the most accurate estimate of $G_{\text {ideal }, B}(\psi)$. To  this end, we formulate the codebook design problem as an MSE as follows,  
\begin{align}
\c_B^{opt} &=\underset{\c, \|\c\|=1}{\arg \min } \int_{-\pi}^{\pi}\left|G_{\text {ideal }, B}(\psi)-G(\psi, \c)\right| d \psi \label{init_opt_eq}
\end{align}

In order to solve the above optimization problem, we rewrite the integral in \eqref{init_opt_eq} as the equivalent infinite series as follows,
\begin{align}
    \c_B^{opt} &= \underset{\c, \|\c\|=1}{\arg \min }\left[\lim _{L \rightarrow \infty} \frac{2\pi}{L}\sum_{\ell=1}^{L} {\left|G_{\text {ideal }, B}\left(\psi_{ \ell}\right)-G\left(\psi_{\ell}, \c\right)\right|}\right]\label{composite_MSE}
\end{align}
where, $\boldsymbol{\psi} = [\psi_1, \cdots, \psi_L]$ is the vector corresponding to the sampled $\psi-\text{domain}$, i.e.,  $\psi_{\ell} = -\pi + \ell(\frac{2\pi}{L})$, $\ell = 1, \cdots, L$. Also, let $\boldsymbol{\psi}_b$ be the elements of $\boldsymbol{\psi}$ that lie in $\omega_b$. It holds that, $|\boldsymbol{\psi}| = L$, and $\psi_b$ is comprised of $|\boldsymbol{\psi}_b| = L_b$, $b \in B$ consecutive elements of $\psi$.  We can rewrite the optimization problem in equation \eqref{composite_MSE} as follows, 
\begin{align}
 &\c_B^{opt} = \underset{\c,  \|\c\|=1}{\arg \min } \lim_{L\rightarrow \infty}\frac{1}{L}|\mathbf{G}_{\text {ideal }, B}-\mathbf{G}(\c)| \label{init_opt}
\end{align}
where,  
\begin{align}
&\mathbf{G}(\c)=\left[ G\left(\psi_{1}, \c\right) \ldots G\left(\psi_{L}, \c\right)\right]^{T} \in \mathbb{Z}^{L}\\
&\mathbf{G}_{\text {ideal }, B}=\left[ G_{\text {ideal }, B}\left(\psi_{1}\right) \ldots G_{\text {ideal }, B}\left(\psi_{ L}\right)\right]^{T} \in \mathbb{Z}^{L}    
\end{align}

Observe that each beam $\mathcal{B}(B)$, divides the angular range into $2|B|$ regions, $|B|$ of which cover the desired ACIs.  Define $\mathbf{e}_{B}(b) \in \mathbb{Z}^{2|B|}$ to be the standard basis vector corresponding to the representation of beam $b$ in the set $\{1, \cdots, 2|B|\}$. For instance, in the example described by fig.~\ref{fig:beam}, we have $e_B([\frac{\pi}{2}, \frac{5\pi}{8})) = [0,0,1,0,0,0]^T$, or $e_B([\frac{-\pi}{4}, \frac{\pi}{8})) = [0,0,0,0,1,0]^T$. Utilizing this notation we write, 
\begin{align}
    \mathbf{G}_{\text {ideal },B}=\sum_{b \in B}\frac{2\pi}{\Delta_B}\left(\mathbf{e}_{B}(b) \otimes \mathbf{1}_{L_b, 1}\right) \label{ideal}
\end{align}

% \amir{either we have to re-index the chosen intervals in $B$ to a number between $1, \ldots, 2|B|$.  Or, we need to define them something like this $\e_B(b)$}

Now, observe that for any equal gain $\mathbf{g} \in \mathbb{C}^{L_b}$ it holds that $\mathbf{1}_{L_b, 1}=\mathbf{g} \odot \mathbf{g}^{*}$ . Therefore, for such choice of $\g$ we can write: 
% $$\mathcal{G}_{L}=\left\{\mathbf{g} \in \mathbb{C}^{L}:\left(\mathbf{g} \mathbf{g}^{H}\right)_{\ell, \ell}=1,1 \leq \ell \leq L\right\}$$
\begin{align}
\mathbf{G}_{\text {ideal }, B} &= \sum_{b \in B}\frac{2\pi}{\Delta_B}\left(\mathbf{e}_{B}(b) \otimes\left(\mathbf{g}_b \odot \mathbf{g}_b^{*}\right)\right) \nonumber\\
&=\sum_{b \in B}\left({\gamma_b}\left(\mathbf{e}_{B}(b) \otimes \mathbf{g}_b\right)\right) \odot\left({{\gamma_b}}\left(\mathbf{e}_{B}(b) \otimes \mathbf{g}_b\right)\right)^{*} \nonumber\\
&=\left(\sum_{b \in B}{\gamma_b}\left(\mathbf{e}_{B}(b) \otimes \mathbf{g}_b\right)\right)  \odot \left(\sum_{b \in B}{\gamma_b}\left(\mathbf{e}_{B}(b) \otimes \mathbf{g}_b\right)\right)^{*} \label{final_gik}
\end{align}

Similarly, one can easily verify that,
\begin{align}
\mathbf{G}(\c) &=\left(\mathbf{D}^{H} \c\right) \odot\left(\mathbf{D}^{H} \c\right)^{*} \label{dc}
\end{align}
where $\mathbf{D}=\left[\mathbf{d}_{M_{t}}\left(\psi_{1}\right) \cdots \mathbf{d}_{M_{t}}\left(\psi_{ L}\right)\right] \in \mathbb{C}^{M_{t} \times L}.$ Given the special form of the equations \eqref{init_opt}, and \eqref{final_gik}, and their usage in the optimization problem entailed in \eqref{dc}, it is straightforward to conclude that $\c^{opt}_B$ is the solution to the following optimization problem for appropriate choices of $\g_b$. 
\begin{problem}
Given any set of equal-gain vectors $\g_b \in \mathbb{C}^{L_b}$, $b \in B$  find vector $\c_B \in \mathbb{C}^{M_t}$ such that
\begin{align}
&\c_B=\underset{\c, \|c\|=1}{\arg \min } \lim_{L\rightarrow \infty} \left\|\sum_{b \in B}{\gamma_b}\left(\mathbf{e}_{B}(b) \otimes \mathbf{g}_b\right)- \mathbf{D}^{H} \c\right\|^{2} \label{obj_func}
\end{align}
\label{main_problem}
\end{problem}
However, in order to find the optimal solution to the optimization problem in \eqref{init_opt}, we need to find the optimal choices of $\g_b$, $b \in B$. Utilizing \eqref{final_gik}, and \eqref{dc}, the following optimization problem arises.
% However, we now need to find the optimal choices for $\g_b$ that minimize the objective in \eqref{init_opt}. Using \eqref{final_gik}, and \eqref{dc}, we have the following optimization problem.
\begin{problem}
Find a set $\mathcal{G}_B$ of equal-gain $\g_b \in \mathbb{C}^{L_b}$,  such that
\begin{equation}
 \mathcal{G}_B = \underset{\mathcal{G}}{\arg\min }\left\| abs(\D^H \c_B)- abs(\sum_{b \in B}{\gamma_b}(\mathbf{e}_{B}(b) \otimes \mathbf{g}_b))\right\|^{2} \label{g_final_eq}
\end{equation} 
where $\mathcal{G}_B = \{\g_b| b \in B \}$, and $abs(.)$ denotes the element-wise absolute value of a vector.
\label{g_problem}
\end{problem}

Therefore, under the fully-digital scheme, by solving problems~\ref{g_problem}, and~\ref{main_problem} the optimal configuration $\c_B^{opt}$ for composite beam $\mathcal{B}(B)$ can be obtained. However, under the hybrid beamforming regime, the optimal configuration is found as
\begin{align}
    \mathbf{F}_{B}, \mathbf{v}_{B} = {\arg\min }_{\mathbf{F},\mathbf{v}} \|\mathbf{F}\mathbf{v} - \c^{opt}_B\|^2
\end{align}
under the equal gain condition on the columns of $\F_B$. Several simple heuristic approaches exist in the literature to obtain near-optimal solutions to the above problem including the effective  orthogonal matching pursuit (OMP)  \cite{love15}\cite{noh17} algorithm. In the next section we continue with the solution to problems~\ref{main_problem}, and~\ref{g_problem}.

\section{Proposed Codebook Design Method}
In this section, we propose our approach for codebook design under the ULA and TULA settings respectively.
\label{sec: proposed}
\subsection{Codebook Design under ULA Setting}
Observe that for each fixed value of $L$ problem~\ref{main_problem} falls in the class of least-square optimization problems. Therefore, as $L$ tends to infinity, $\c_B$ is obtained as the limit of the solutions to this problem. For each $L$ the solution is given by,
 \begin{align}
& {\c}^{(L)}_B = \sum_{b \in B}{\gamma_b}(\D \D^H)^{-1} \D  \left(\mathbf{e}_{B}(b) \otimes \mathbf{g}_b\right) \\
& {\c}^{(L)}_B = \frac{1}{L}\sum_{b \in B}\gamma_b \D( \mathbf{e}_{B}(b) \otimes \g_b) \label{c_final_eq}
\end{align}
% where $\sigma_s = \frac{\gamma_s}{L} = \frac{\sqrt{\frac{2\pi}{\Delta_B}}}{L}$, 
where it holds that, 
$$\mathbf{D D}^{H}  = \sum_{l=1}^L \mathbf{d}_{M_{t}}\left(\psi_{l}\right)\mathbf{d}^H_{M_{t}}\left(\psi_{l}\right) = \sum_{l=1}^L \I_{M_t} = L\I_{M_t}$$
% $$\mathbf{D D}^{H}=\left({L \sum_{p=1}^{2^B}\delta_p}\right) \mathbf{I}_{M_{t}}$$.
% % Dividing by $\|\Tilde{\c}^{(L)}_q\|$ and taking the limit as L goes to infinity we will find the optimal $\c_q$. i.e. 
Define $\boldsymbol{\Gamma}_\mathrm{B} = \sum_{b \in B}{\gamma_b}(\mathbf{e}_{B}(b) \otimes \mathbf{g}_b))$. Replacing $\boldsymbol{\Gamma}_\mathrm{B}$ and \eqref{c_final_eq} in equation \eqref{g_final_eq} we get, 
% \begin{align}
%  \mathcal{G}_B  = \underset{\mathcal{G}}{\arg\min }& \left\| abs(\D^H \D\sum_{b \in B}\frac{\gamma_b}{L}(\mathbf{e}_{B}(b) \otimes \mathbf{g}_b))\right . \nonumber \\
%  &\left . - abs(\sum_{b \in B}{\gamma_b}(\mathbf{e}_{B}(b) \otimes \mathbf{g}_b))\right\|^2 \label{g_simple_final_eq}
% \end{align}
\begin{align}
 \mathcal{G}_B  = \underset{\mathcal{G}}{\arg\min }& \left\| abs(\frac{\D^H\D}{L}\boldsymbol{\Gamma}_\mathrm{B}) - abs(\boldsymbol{\Gamma}_\mathrm{B})\right\|^2. \label{g_simple_final_eq}
\end{align}
%where $\sigma' = \frac{1}{2\pi L}$. 
% We will use the following proposition without providing formal proof due to space limitation.

% \begin{proposition}
% The minimizer of \eqref{g_simple_final_eq} is in the form of geometric vectors $\g_b$, of lengths $L_b$, and parameters $\rho_b = \frac{\eta\delta_b}{L}$, $b \in B$, for some real value $\eta\geq 0$. \label{proposiiton_g}
% \end{proposition}

Even though Problem~\ref{main_problem} admits a nice analytical closed form solution, doing so for the Problem~\ref{g_problem} is not a trivial task, especially due to the fact that the objective function is not convex. However, the convexification of \eqref{g_final_eq} in the form of 
\begin{align}
 & \mathcal{G}_{\mathrm{B}}=
{\arg\min }_{\mathcal{G}}\left\| \left(\frac{1}{L}\D^H \D- \I_{LQ}\right) \boldsymbol{\Gamma}_{\mathrm{B}}\right\|^{2} \label{g_final_eq_convexified}
\end{align} 
leads to an effective solution for the original problem.
Indeed, it can be verified by solving the optimization problem \eqref{g_final_eq_convexified} numerically that a close-to-optimal solution admits the form of geometric vectors $\g_b$, of lengths $L_b$, and parameters $\rho_b = \frac{\eta\delta_b}{L}$, $b \in B$, for some real value $\eta$.
%Indeed, it can be verified by solving the optimization problem \eqref{g_final_eq_convexified} numerically that the solution is in the form \eqref{g_conjecture} which leads us to the following conjecture. 
% \begin{conjecture}
% The minimizer of \eqref{g_final_eq_convexified} is in the form of 
% \begin{align}
%     \g^*_{p,q} = \left[{ 1,  \alpha_v \alpha_h, \cdots,  \alpha_v^{ (L_v -1)}\alpha_h^{ (L_h -1)} }\right]^T, (p,q) \in \mathcal{A} \label{g_conjecture}
% \end{align}
% for some $\eta_v$, $\eta_h$  where $\alpha_a = e^{j(\frac{\eta_a}{L_a})}$, $a \in \{v, h\}$. \label{proposiiton_g}
% \end{conjecture}
%
% Except for some special cases, we have not been able to analytically prove this conjecture in its entirety. 
We use this analytical form for $\g_{b}$ for the rest of our derivations. This solution would not be the optimal solution for the original problem \eqref{g_final_eq}.  However, it provides a near optimal solution with added benefits of allowing to (i) find the limit of the solution as $L$ goes to infinity, (ii) express the beamforming vectors in closed form, as it will be revealed in the following discussion, and (iii) change $\eta$ in order to design beams with different qualities such as beamforming gain, smoothness, and leakage as it is explained in more details in section~\ref{sec:evaluation}.
%\amir{ find the proof.}
Let us expand the expression for ${\c}_{B}^{(L)}$ from equation \eqref{c_final_eq} to get,
 \begin{align}
     {\c_B}^{(L)} & = \sum_{b \in B} \frac{\gamma_b}{L} \left(\sum_{\ell=1}^{L_b}g_{b,\ell}\mathbf{d}_{M_t}(\psi_{b, \ell})\right)  
    %  \nonumber\\
    %  & = \sum_{b \in B} \left(\frac{\gamma_b}{L}\sum_{l=1}^{L_b} g_{b,l}\left[{\begin{array}{ccc}
    %  1 &\cdots & e^{j(M_t-1)\psi_{b, l}}\\
    %  \end{array}}\right]^T \right)  
    %  & = \sum_{q \in \mathcal{W}_k} \sigma_q \left[{\begin{array}{ccc}
    %  \sum_{l=1}^L g_{q, l}& \cdots & \sum_{l=1}^L g_{q, l}e^{j(M_t-1)\psi_{q, l}}\\
    %  \end{array}}\right]^T 
\end{align}

% Let us write, 

% \begin{equation}
%     \c_k^{(L)} = \sum_{q \in \mathcal{W}_k} \c_q^{(L)}
% \end{equation}

Further define, 
\begin{align}
\c_b^{(L)}  = \frac{\gamma_b}{L} \sum_{\ell=1}^{L_b}g_{b, \ell}\mathbf{d}_{M_t}(\psi_{b, \ell}), \quad b \in \mathrm{B}
\end{align}

% to get,

% \begin{equation}
%     \c_B^{(L)} = \sum_{b \in B} \c_b^{(L)}
% \end{equation}

The $m$-th element of the vector ${\c}_b = \lim_{L\rightarrow \infty}{\c_b}^{(L)} $, i.e. $c_{b,m}$, is given by 
\begin{align}
    c_{b,m} &= \sqrt{\frac{2\pi}{\Delta_B}}\lim_{L\rightarrow \infty} \frac{1}{L}\sum_{\ell=0}^{L_b-1} g_{b, \ell} e^{jm\psi_{b, \ell}} \nonumber\\
    & = \frac{\delta_b}{\sqrt{ 2\pi\Delta_B}}\lim_{L_b\rightarrow \infty} \frac{1}{L_b}\sum_{\ell=0}^{L_b-1} g_{b, \ell} e^{jm\psi_{b, \ell}}
\end{align}

% With a choice of $\g_q = \left[{\begin{array}{cccc} 1& \alpha^\eta &\cdots & \alpha^{\eta (L -1)} \end{array}}\right]^T$ where we set 
% % $\alpha = e^j(\frac{2\pi}{L2^B})$
% $\alpha = e^{j(\frac{\delta_q}{L})}$ and suitable $\eta$ (to be determined later), we can write

% \begin{equation}
%     \Tilde{c}_q^{(m)} = \lim_{L\longrightarrow \infty} \frac{1}{L}\sum_{l=0}^{L-1} g_l e^{jm(\psi_{q-1}+\frac{2 \pi(\ell+0.5)}{L2^B})}
% \end{equation}

For large enough $L$, we can write $\psi_{\ell, b} = \psi_b^s + \ell(\frac{\delta_b}{L_b})$ to get
% Choosing $\g_b$ as in proposition \ref{proposiiton_g} we have %and replacing $\psi_{b,l} = \psi_b^s + l\frac{\delta_b}{L_b}$
\begin{align}
    c_{b,m} &= \frac{\delta_b}{\sqrt{ 2\pi\Delta_B}}\lim_{L_b\rightarrow \infty} \frac{1}{L_b}\sum_{\ell=0}^{L_b-1} e^{j\frac{\ell\eta\delta_b}{L_b}} e^{jm(\psi_b^s + \ell\frac{\delta_b}{L_b})} \nonumber \\
    &= \frac{\delta_b}{\sqrt{ 2\pi\Delta_B}}e^{jm(\psi^s_{b})}\lim_{L\rightarrow \infty} \frac{1}{L_b}\sum_{\ell=0}^{L_b-1} \alpha^{(\eta+m) \ell}   \nonumber \\
    &= \frac{\delta_b}{\sqrt{ 2\pi\Delta_B}}e^{jm(\psi^s_{b})} \int_{0}^{1} \alpha^{(\eta + m)L_bx}dx
\end{align}

\noindent where $\alpha = e^{j\frac{\delta_b}{L_b}}$. After a few straightforward steps, we get,  
% \begin{equation}
%     c_{q,m} = \frac{\delta_q}{\sqrt{2\pi}\Delta_k}\lim_{L\rightarrow \infty} \frac{1}{L}\sum_{l=0}^{L-1} g_{q, l} e^{jm(\psi_{q-1}+\frac{\delta_q(\ell+0.5)}{L})}
% \end{equation}
% % After some basic manipulations we get, 
% \begin{equation}
%     c_{q,m} = \frac{\delta_q}{\sqrt{2\pi}\Delta_k}\lim_{L\rightarrow \infty} \frac{1}{L}\sum_{l=0}^{L-1} \alpha^{(\eta+m) l} e^{jm(\psi_{q-1}+\frac{0.5\delta_q}{L})} 
% \end{equation}
% \begin{equation}
%     c_{b,m} = \frac{\delta_b}{\sqrt{ 2\pi\Delta_B}}e^{jm(\psi^s_{b})}\lim_{L\rightarrow \infty} \frac{1}{L_b}\sum_{l=0}^{L_b-1} \alpha^{(\eta+m) l}   
% \end{equation}
% \begin{equation}
%     c_{b,m} = \frac{\delta_b}{\sqrt{ 2\pi\Delta_B}}e^{jm(\psi^s_{b})} \int_{0}^{1} \alpha^{(\eta + m)L_bx}dx
% \end{equation}
% \begin{equation}
%     c_{q,m} = \frac{\delta_q}{\sqrt{2\pi}\Delta_k}e^{jm(\psi_{q-1})} \int_{0}^{1} e^{j\frac{2\pi(\eta + m)L}{L2^B}x}dx
% \end{equation}
\begin{align}
 c_{b,m}= \frac{\delta_b}{\sqrt{ 2\pi\Delta_B}}e^{j(m\psi^s_{b} + \frac{\xi}{2})} sinc(\frac{\xi}{2\pi}) \label{final_g}
\end{align}
where $\xi = \delta_b (\eta +m )$.
%\nariman{Here, we can motive the use of tula by highlighting the inefficiencies especially that the formation of a composite beam gets faulty + the two points you have mentioned in the evaluation part + citation from the MILCOM paper.}
We note that the ULA antenna structure inherently generates two-sided lobes and hence inefficient beams due to (i) having beam lobes in undesired scopes, and (ii) having lower effective beam gain in desired ACIs. In the following we discuss a beam design using TULA antenna structure. The TULA structure not only generates single sided beams, but also improves the beam gain (by almost 3 dB)\cite{TULA}.

\subsection{Codebook Design under TULA Setting}

For some $\beta\in \mathbb{R}$ define the codeword under the TULA configuration as, 
\begin{align}
    & \c_{B, twin} = \left[\c^T_{B, twin, \frac{M_t}{2}}, e^{j\beta}\c^T_{B, twin, \frac{M_t}{2}}\right]^T \label{new_c_twin}
\end{align}

Under such codeword, appealing to \eqref{tula_gain} and \eqref{tula_dir}, we can rewrite the corresponding reference gain as, 
\begin{align}
    G(\theta, \c_{B, twin}) &= L(\theta)^2 \left|\d_{\frac{M_{t}}{2}}^{H}(\theta)\c_{B, twin, \frac{M_t}{2}}\right|^2
    % & = \left|\d_{\frac{M_{t}}{2}}^{H}(\theta)\c\right|^2 \left|{1 + e^{j(\beta-\frac{2\pi}{3}\sin(\theta))}}\right|^2
    \label{final_gain}
\end{align}

where we define, 
\begin{align}
    L(\theta) = \left| {1 + e^{j(\beta-\frac{2\pi}{3}\sin(\theta))}}\right| = \left|\cos(\frac{\beta}{2}-\frac{\pi}{3}\sin(\theta))\right|
\end{align}

We note that the gain in \eqref{final_gain} consists of two terms, one being the reference gain of a ULA of $M_t/2$ antennas (by setting the same codeword as in \eqref{final_g}) and one being a function of the parameter $\beta$. As before, we use the first term to obtain the beamforming gain, but use the second term to provide the required level of isolation between each beam and its mirrored counterpart to resolve the inefficiency of the ULA structure. 
% More precisely, based on equation \eqref{final_g}, we set 
% \begin{align}
%  &c_{B, twin, m} = \sum_{b \in B}c_{b, twin, m} \nonumber\\
%  &= \sum_{b \in B} \frac{\delta_b}{\sqrt{ 2\pi\Delta_B}}e^{j(m\psi^s_{b} + \frac{\xi}{2})} sinc(\frac{\xi}{2\pi}) , \quad m = 0 \ldots \frac{M_t}{2} -1
%     \label{twin_c_result}
% \end{align}
To capture the isolation requirement, we define the \emph{isolation factor} $ 0 \leq \mu <1$ as follows
\begin{align}
    \mu = \underset{\omega_b}{\int}{\frac{L(-\theta)}{L(\theta)}} d\theta = \underset{\omega_b}{\int}{\frac{\cos(\frac{\beta}{2}+\frac{\pi}{3}\sin(\theta))}{\cos(\frac{\beta}{2}-\frac{\pi}{3}\sin(\theta))}} d\theta \label{isolation}
\end{align}
 to denote the level of isolation between each $\omega_b$ and its mirrored counterpart. Opting for a small enough value for $\mu$, optimal $\beta$ has to be obtained by numerically solving equation \eqref{isolation}. Plugging in the obtained $\beta$ into equation \eqref{new_c_twin} completes the codebook design under the TULA structure.

\section{Performance Evaluation}
\label{sec:evaluation}

\begin{figure*}
    \centering
    \begin{minipage}[h]{0.25\textwidth}
        \includegraphics[width=1\linewidth]{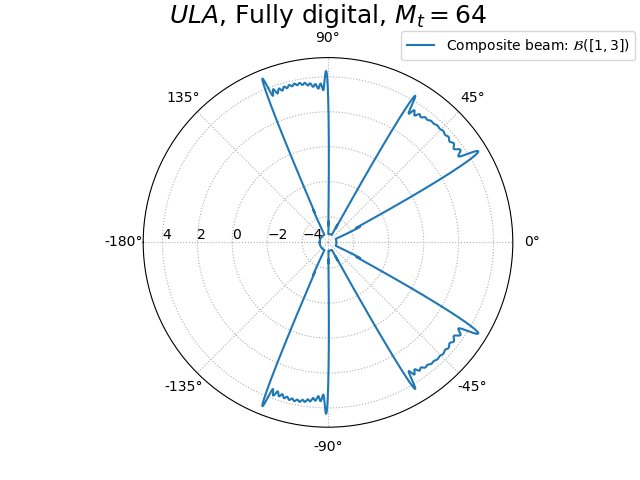}
        \caption{  Fully-digital, ULA }
        \label{fig:ULA_comp}
    \end{minipage}
    \begin{minipage}[h]{0.24\textwidth}
        \includegraphics[width=1\linewidth]{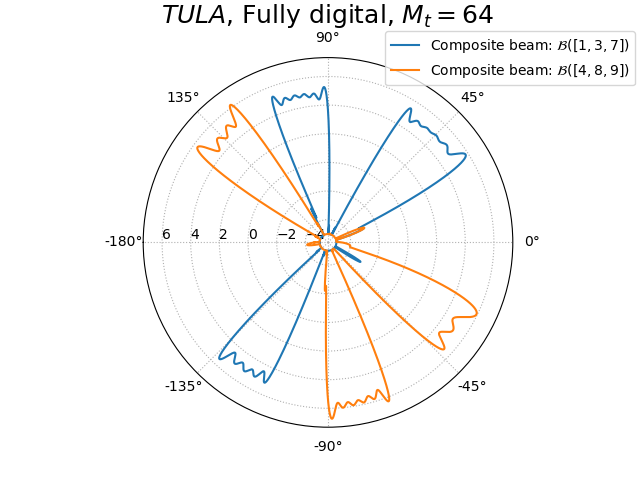}
        \caption{  Fully-digital, TULA}
        \label{fig:TULA_comp}
    \end{minipage}
    \begin{minipage}[h]{0.24\textwidth}
        \includegraphics[width=1\linewidth]{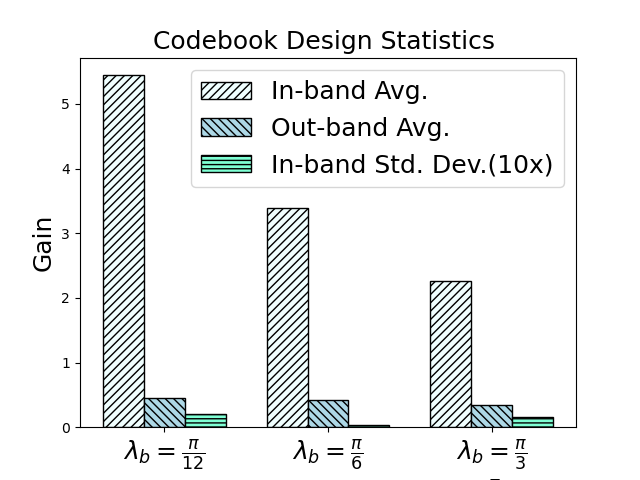}
        \caption{ Beam quality vs. $\lambda_b$}
        \label{fig:stat_beam}
    \end{minipage}
    \begin{minipage}[h]{0.24\textwidth}
        \includegraphics[width=1\linewidth]{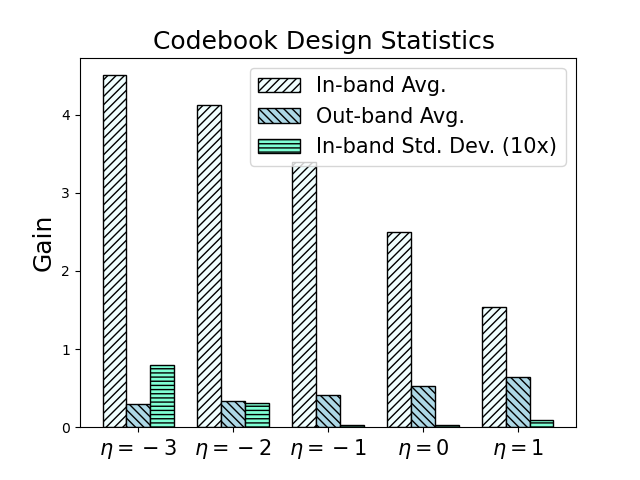}
        \caption{ Beam quality vs. $\eta$}
        \label{fig:stat_eta}
    \end{minipage}
\end{figure*}

In this section, we provide performance evaluation of our proposed composite beamforming technique using numerical analysis. 
%We consider ULA and TULA structures for the array of antennas and  design the composite beams from \emph{Example $1$} that are depicted in Fig.~\ref{fig:example}. 
We consider ACI set and corresponding indices as in Example~1.
Fig.~\ref{fig:ULA_comp} depicts the composite beam corresponding to $\mathcal{B}(\{1,3\})$ designed for ULA, where the beam gain is depicted in dB. It is observed that the beams have smooth gains within the desired ACIs with very sharp edges and negligible out-of-band leakage. 
%\nariman{Maybe we can move the following to points to the end o spart A of the last section when we want to motivate tula}
%However, the ULA structure inherently generates two-sided lobes and hence generates inefficient beams due to (i) having beam lobes in undesired scopes, and (ii) having lower effective beam gain in desired ACIs (by almost 3 dB). 
%Moreover, ULA is incapable of forming composite beams such as $\mathcal{B}(B_1)$, and $\mathcal{B}(B_2)$, i.e. beams with some lobes in $[-\pi, 0]$, and some lobes $[0, \pi]$. 
%To deal with this issue, we exploit the TULA antennas structure. 
As illustrated in Fig.~\ref{fig:TULA_comp}, by using TULA antenna structure, our beam design technique is capable of covering beams $\mathcal{B}(B_1)$, and $\mathcal{B}(B_2)$ as in Example~1 with high stable gain, while resolving the two-sided beam issue arises in ULA. 

\begin{figure*}
    \centering
            \subfloat[Single-beam shape, $\eta = -2$ \label{eta-2}]{%4
            \includegraphics[width=0.25\linewidth]{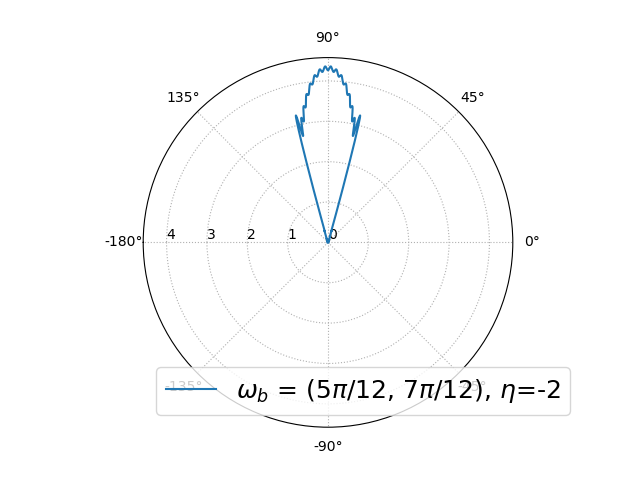}}
            % \subfloat[One-sided TULA \label{one-sided64}]{%4
            % \includegraphics[width=0.225\linewidth]{figures/single_beam -1.5.png}}
            \subfloat[Single-beam shape, $\eta = -1$ \label{eta-1}]{%4
            \includegraphics[width=0.25\linewidth]{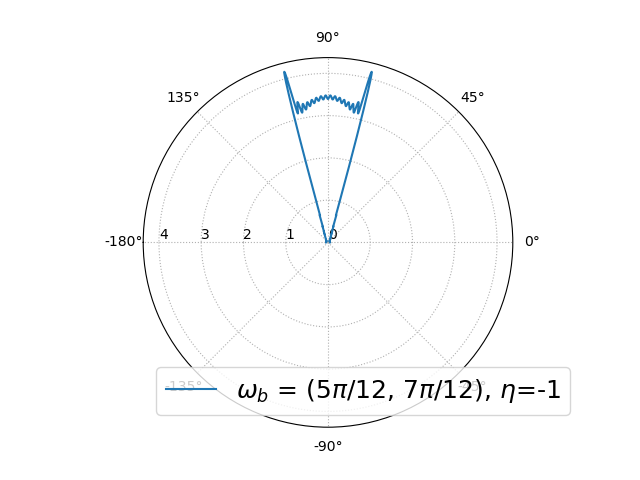}}
            % \subfloat[One-sided TULA \label{one-sided64}]{%4
            % \includegraphics[width=0.225\linewidth]{figures/single_beam -0.75.png}}
            % \subfloat[One-sided TULA \label{one-sided64}]{%4
            % \includegraphics[width=0.225\linewidth]{figures/single_beam -0.5.png}}
            \subfloat[Single-beam shape, $\eta = 0$ \label{eta0}]{%4
            \includegraphics[width=0.25\linewidth]{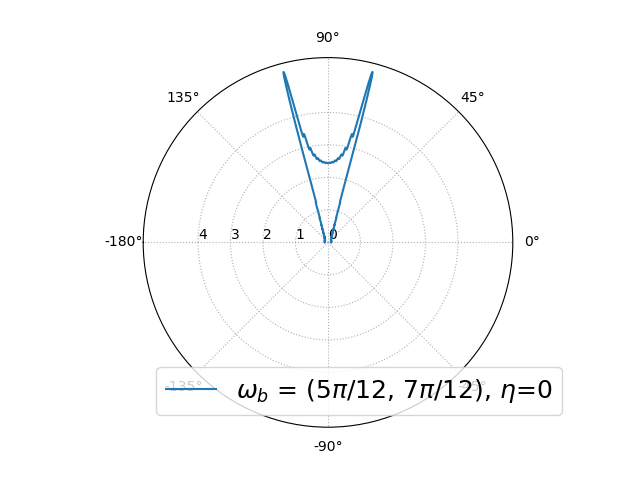}}
            \subfloat[Single-beam shape, $\eta = 1$ \label{eta1}]{%4
            \includegraphics[width=0.25\linewidth]{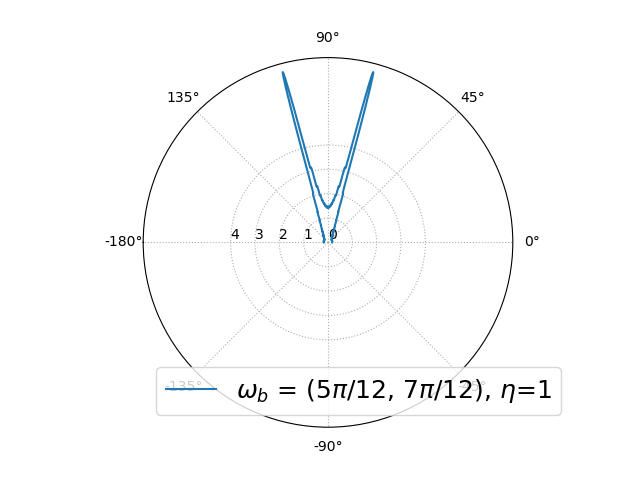}}
            
            \caption{Single-beam shape for varying $\eta$}
            \label{single_eta}
\end{figure*}

In order to quantify three main qualities of a beam, i.e., the gain, leakage, and smoothness we define three performance metrics, namely the \emph{in-band average gain} of the beam, the \emph{out-band average gain}, and its \emph{in-band variance}, respectively. Fig.~\ref{fig:stat_beam} presents the evaluation of these metrics for a single beam centered around $\theta = \frac{\pi}{2}$ versus its beamwidth. Note that, the amount of in-band variance and out-band average gain are negligible, which confirms the high quality of the beams generated by our design. Moreover, as intuition also suggests, it is observed that the beam gain is almost inversely proportional to its beamwidth. %increases the in-band gain decreases linearly, that is in-line with our theoretical findings for the in-band gain. 
Another important design parameter is $\eta$ introduced in the definition of the geometric vector form for $\g$. Fig.~\ref{fig:stat_eta} shows how varying the value of $\eta$ impacts the beam quality measures, for a single-beam centered around $\theta = \frac{\pi}{2}$, with $\lambda_b = \frac{\pi}{6}$. Of course, based on the design parameter $\eta$, there is a three-way trade off between the smoothness, gain and leakage. For the rest of this section we have used $\eta = -1$ which results in the smoothest beam gain with an acceptable in-band gain.% is picked for the numerical simulations. 
%It is observed that $\eta = -1$ results in the smoothest beam gain with an acceptable in-band gain. 
In order to provide better envisioning of the beam shape for the reader, Fig.~\ref{single_eta} shows how changing the value of $\eta$ effects the beam pattern for a single beam.

\begin{figure*}
    \centering
    
    \subfloat[Hybrid \label{hyb}]{%4
            \includegraphics[width=0.25\linewidth]{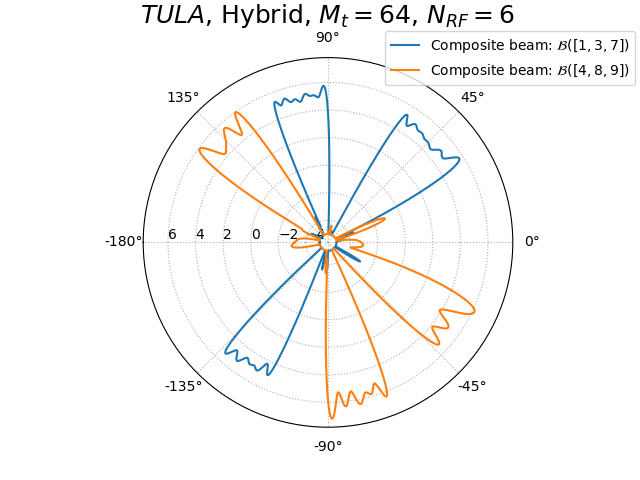}}
            % \subfloat[One-sided TULA \label{one-sided64}]{%4
            % \includegraphics[width=0.225\linewidth]{figures/single_beam -1.5.png}}
            \subfloat[Hybrid-Quantized \label{hyb-q}]{%4
            \includegraphics[width=0.25\linewidth]{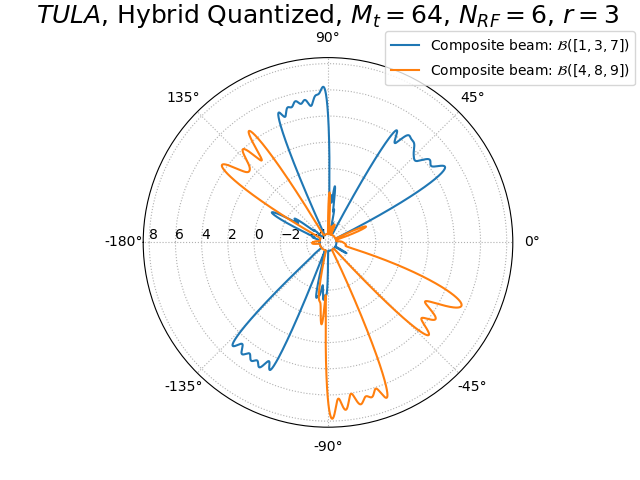}}
            % \subfloat[One-sided TULA \label{one-sided64}]{%4
            % \includegraphics[width=0.225\linewidth]{figures/single_beam -0.75.png}}
            % \subfloat[One-sided TULA \label{one-sided64}]{%4
            % \includegraphics[width=0.225\linewidth]{figures/single_beam -0.5.png}}
            \subfloat[Quantized-Hybrid \label{q-hyb}]{%4
            \includegraphics[width=0.25\linewidth]{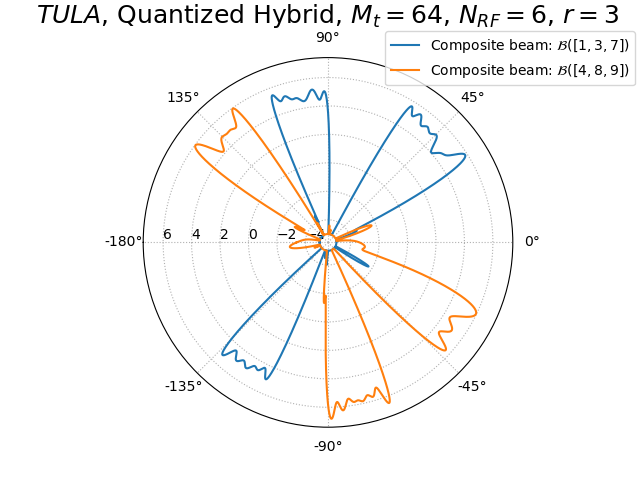}}
            \subfloat[Beam quality vs. $r$ \label{hq_stat}]{%4
            \includegraphics[width=0.25\linewidth]{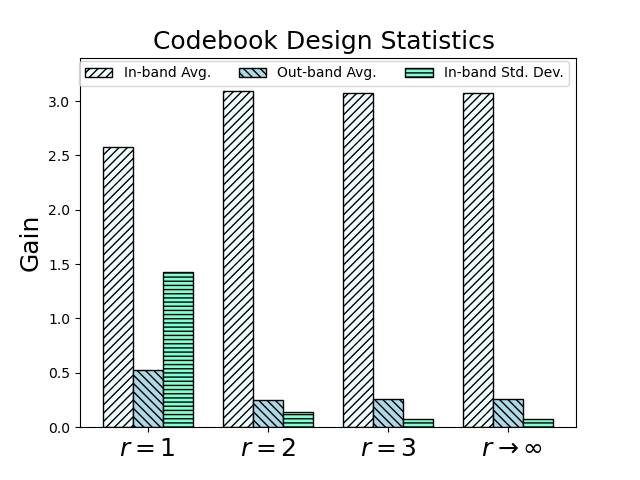}}
            \caption{Effect of quantization on hybrid beamforming using TULA}
            \label{fig:TULA_HYB_COMP}
\end{figure*}

Figures~\ref{fig:ULA_comp}-\ref{single_eta} illustrates the simulation results for a system which is capable of fully-digital beamforming, i.e., where each antenna element is wired to a single RF chain. However, to reduce the number of RF chains hybrid beamforming is employed, where only a few (say $6$) RF chains control a second layer phased array feed line to the antennas. Hence a beamformer $\c$ is approximated by $\F\v$ where column of matrix $\F$ correspond to the phase only elements. Although suboptimal, an efficient algorithm to find $\F$ from $\c$ is OMP. 
Fig.~\ref{hyb} shows the result of running the OMP algorithm where $N_{RF} =6$ RF  chains are used. However, we note that the codebook has to be stored with finite resolution, say $r$ bits, for each phase shift corresponding to the entries of $\F$. 
%Suppose the resolution of quantization of the phase is denoted by $r$ bits, i.e. $2^r$ different values for phase shifts are available for analog beamforming stage of the hybrid beamforming sysgem. 
%There is a trade-off between the complexity of the beamformer and the accuracy of the phase shifts, i.e. larger $r$ results in more accurate estimate of the phase shifts at the cost of more complex beamformers. 
A naive approach is to quantize the entries of $\F$ obtained using OMP using $r$ bits and the result of such approach for $r=3$ is given in Fig.~\ref{hyb-q}. Comparing Fig.~\ref{hyb}~and~\ref{hyb-q}, it is observed that such inevitable quantization considerably affects the beam shape and beam gain, hence it suggests that we require finer quantizations.
%is obtained when the optimal phase shifts resulting from the OMP algorithm are mapped to the closest quantized level, when $r=3$. 
%Further, to increase the accuracy of the quantized phase shifts, we propose to integrate the quantization step with the OMP algorithm, such that at each iteration the resulting phase shift vector gets quantized with $r=3$ and then the next iteration follows. 
However, a more elaborate technique can be used where the quantization is performed in each steps of the OMP algorithm. The difference is in the former approach we find the hybrid beamforming matrix $\F$ and then quantize it, i.e., hybrid-quantized, while in the latter approch we perform quantization at each step, i.e., quantized-hybrid. Using quantized-hybrid approach, Fig.~\ref{q-hyb} shows that the beam pattern and its gain for $r=3$ is very similar to that of Fig.~\ref{hyb} where there is no limit on the resolution, i.e., $r=\infty$. Finally, Fig.~\ref{hq_stat} shows the effect of increasing the quantization resolution in the quality of the beams generated by the quantized-hybrid approach. It is observed that at $r=3$, the quality of the generated beams levels that of the hybrid scheme, denoted by $r\rightarrow \infty$. 
%\amir{should we remove $r=4$ since it is better than $r=\infty$. it is hard to explain why, etc.}

\section{Conclusions}
\label{sec:conclusions}

We studied the composite codebook design problem and illustrated how multiple disjoint ACIs with different beamwidths can be covered with a single codeword. We highlighted the inefficiencies of ULA in forming arbitrary composite beams and showed that how we can overcome these inefficiencies by employing a novel antenna structure, namely TULA. We derived a low-complexity analytical closed-form solution for the composite codebook design problem for both the ULA and the TULA case and confirmed the validity of our theoretical findings by means of numerical experiments. 

% \input{derivation}
% \input{L_free_formulation.tex}

% The authors would like to thank...

% trigger a \newpage just before the given reference
% number - used to balance the columns on the last page
% adjust value as needed - may need to be readjusted if
% the document is modified later
%\IEEEtriggeratref{8}
% The "triggered" command can be changed if desired:
%\IEEEtriggercmd{\enlargethispage{-5in}}

% references section

% can use a bibliography generated by BibTeX as a .bbl file
% BibTeX documentation can be easily obtained at:
% http://www.ctan.org/tex-archive/biblio/bibtex/contrib/doc/
% The IEEEtran BibTeX style support page is at:
% http://www.michaelshell.org/tex/ieeetran/bibtex/
%\bibliographystyle{IEEEtran}
% argument is your BibTeX string definitions and bibliography database(s)
%\bibliography{IEEEabrv,../bib/paper}
%
% <OR> manually copy in the resultant .bbl file
% set second argument of \begin to the number of references
% (used to reserve space for the reference number labels box)

\bibliographystyle{IEEEtran}
\bibliography{bibliography}

% Generated by IEEEtran.bst, version: 1.14 (2015/08/26)
\begin{thebibliography}{10}
\providecommand{\url}[1]{#1}
\csname url@samestyle\endcsname
\providecommand{\newblock}{\relax}
\providecommand{\bibinfo}[2]{#2}
\providecommand{\BIBentrySTDinterwordspacing}{\spaceskip=0pt\relax}
\providecommand{\BIBentryALTinterwordstretchfactor}{4}
\providecommand{\BIBentryALTinterwordspacing}{\spaceskip=\fontdimen2\font plus
\BIBentryALTinterwordstretchfactor\fontdimen3\font minus
  \fontdimen4\font\relax}
\providecommand{\BIBforeignlanguage}[2]{{%
\expandafter\ifx\csname l@#1\endcsname\relax
\typeout{** WARNING: IEEEtran.bst: No hyphenation pattern has been}%
\typeout{** loaded for the language `#1'. Using the pattern for}%
\typeout{** the default language instead.}%
\else
\language=\csname l@#1\endcsname
\fi
#2}}
\providecommand{\BIBdecl}{\relax}
\BIBdecl

\bibitem{nitsche2015steering}
T.~Nitsche, A.~Flores, E.~Knightly, and J.~Widmer, ``Steering with eyes closed:
  mm-wave beam steering without in-band measurement,'' 2015.

\bibitem{noh17}
S.~Noh, M.~D. Zoltowski, and D.~J. Love, ``Multi-resolution codebook and
  adaptive beamforming sequence design for millimeter wave beam alignment,''
  \emph{IEEE Transactions on Wireless Communications}, vol.~16, no.~9, pp.
  5689--5701, 2017.

\bibitem{khal20}
M.~Hussain and N.~Michelusi, ``Energy-efficient interactive beam alignment for
  millimeter-wave networks,'' \emph{IEEE Transactions on Wireless
  Communications}, vol.~18, no.~2, pp. 838--851, 2019.

\bibitem{SS19}
C.~N. Barati, S.~A. Hosseini, M.~Mezzavilla, T.~Korakis, S.~S. Panwar,
  S.~Rangan, and M.~Zorzi, ``Initial access in millimeter wave cellular
  systems,'' \emph{IEEE Transactions on Wireless Communications}, vol.~15,
  no.~12, pp. 7926--7940, 2016.

\bibitem{shah20}
M.~A. Amir~Khojastepour, S.~Shahsavari, A.~Khalili, and E.~Erkip, ``Multi-user
  beam alignment for millimeter wave systems in multi-path environments,'' in
  \emph{2020 54th Asilomar Conference on Signals, Systems, and Computers},
  2020, pp. 549--553.

\bibitem{shah19}
N.~Michelusi and M.~Hussain, ``Optimal beam-sweeping and communication in
  mobile millimeter-wave networks,'' in \emph{2018 IEEE International
  Conference on Communications (ICC)}, 2018, pp. 1--6.

\bibitem{Nos19}
M.~Nosrati, S.~Shahsavari, S.~Lee, H.~Wang, and N.~Tavassolian, ``A concurrent
  dual-beam phased-array doppler radar using mimo beamforming techniques for
  short-range vital-signs monitoring,'' \emph{IEEE Transactions on Antennas and
  Propagation}, vol.~67, no.~4, pp. 2390--2404, 2019.

\bibitem{Ata20}
S.~Atapattu, R.~Fan, P.~Dharmawansa, G.~Wang, J.~Evans, and T.~A. Tsiftsis,
  ``Reconfigurable intelligent surface assisted two–way communications:
  Performance analysis and optimization,'' \emph{IEEE Transactions on
  Communications}, vol.~68, no.~10, pp. 6552--6567, 2020.

\bibitem{glo21}
\BIBentryALTinterwordspacing
N.~{Torkzaban} and M.~{Khojastepour}, ``Shaping mmwave wireless channel via
  multi-beam design using reconfigurable intelligent surfaces,'' in \emph{2021
  IEEE Global Communications Conference- GLOBECOM (Accepted)}, 2021. [Online].
  Available:
  \url{{https://www.nec-labs.com/uploads/Documents/Mobile-Communications/Shaping_mmWave_Wireless_Channel_via_Multi_BeamDesign_using_Reconfigurable_Intelligent_Surfaces_Globecom_.pdf}}
\BIBentrySTDinterwordspacing

\bibitem{gholami2020joint}
A.~Gholami, N.~Torkzaban, J.~Baras, and C.~Papagianni, ``Joint mobility-aware
  uav placement and routing in multi-hop uav relaying systems,'' in \emph{Ad
  Hoc Networks}.\hskip 1em plus 0.5em minus 0.4em\relax Cham: Springer Intl.
  Pub., 2021, pp. 55--69.

\bibitem{love15}
J.~Song, J.~Choi, and D.~J. Love, ``Codebook design for hybrid beamforming in
  millimeter wave systems,'' in \emph{2015 IEEE International Conference on
  Communications (ICC)}, 2015, pp. 1298--1303.

\bibitem{TULA}
\BIBentryALTinterwordspacing
N.~{Torkzaban} and M.~{Khojastepour}, ``Codebook design for beamforming in 5g
  and beyond mmwave systems,'' in \emph{2022 IEEE ICC (Accepted)}, 2021.
  [Online]. Available:
  \url{{https://www.nec-labs.com/uploads/Documents/Mobile-Communications/Codebook_Design_for_Single_Beam_MILCOM.pdf}}
\BIBentrySTDinterwordspacing

\end{thebibliography}

\end{document}